\documentclass[12pt]{article}
\usepackage{amssymb}

\newcommand{\sect}[1]{\setcounter{equation}{0}\section{#1}}

\def\be{\begin{equation}}
\def\ee{\end{equation}}
\def\ba{\begin{eqnarray}}
\def\ea{\end{eqnarray}}

\title{\vspace{-1in}
\parbox{\linewidth}
{\small\hfill HUTP-01/A050}\\
{\small\hfill ITEP-TH-59/01}\\
{\small\hfill RUNHETC-2002-01}\\
{\small\hfill hep-th/0201227}\\
\vspace{0.6in}
{\bf $G$-Flux, Supersymmetry and $Spin(7)$ Manifolds}}
\author{
\textsc{Bobby S. Acharya}\\
\emph{Department of Physics, Rutgers University}\\
\emph{Piscataway, NJ 08854, USA}\and
\textsc{Xenia De La Ossa}\\
\emph{Mathematical Institute, University of Oxford,} \\
\emph{24-29 St Giles', Oxford. OX1 3LB. UK}\and
\textsc{Sergei Gukov}\\
\emph{Jefferson Physical Laboratory, Harvard University}\\
\emph{Cambridge, MA 02138, USA}}

\begin{document}
\pagestyle{plain}
\setcounter{page}{1}
\newcounter{bean}
\baselineskip16pt

\maketitle

\begin{abstract}

In this note we study warped compactifications of $M$-theory
on manifolds of $Spin(7)$ holonomy in the presence of background
4-form flux.
The explicit expression for the superpotential can be given
in terms of the self-dual Cayley calibration on the $Spin(7)$ manifold,
in agreement with the general formula proposed in hep-th/9911011.

\end{abstract}

\vspace{0.5in}

\newpage


\sect{Introduction}

Various aspects of $M$-theory compactifications on manifolds of exceptional
holonomy and related vacua have been studied recently $[1-46]$.
This is partly due to their relation to
minimally supersymmetric gauge theories.
Although $Spin(7)$ manifolds are perhaps less relevant to the construction
of realistic models than $G_2$ manifolds, they expose other aspects
of $M$-theory, related to the interesting dynamics of the $\mathcal{N}=1$
effective theory in $2+1$ dimensions \cite{GS}.
In compactification on $G_2$-manifolds supersymmetry
and zero cosmological constant require the 4-form field
strength $G$ to vanish \cite{CR,DH,AS},
whereas -- as we will see presently -- in $M$-theory
on $Spin(7)$-manifolds there is more freedom,
and non-trivial $G$-flux can be consistent with
supersymmetry.

In fact, there are several reasons why non-trivial
$G$-flux may be required in $M$-theory compactifications
on $Spin(7)$-manifolds.
For example, cancellation of membrane
anomalies in an arbitrary vacuum spacetime
forces the 4-form flux $G$ to obey
the modified quantization condition \cite{witten1}:
\be
\Big[ {G \over 2 \pi} \Big] - {\lambda \over 2}
\in H^4 (X, \mathbb{Z})
\label{qcond}
\ee
where
the integral class $\lambda=p_1(X)/2\in{H}^4(X;\mathbb{Z})$
and $X$ is the compactification manifold.
If $\lambda$ is even, then $G=0$ is a consistent part of the
vacuum data.
In particular, in $M$-theory on $G_2$ holonomy manifolds,
if the anomaly did not vanish
the corresponding
compactifications would not lead to supersymmetric
vacua because $G$ would have to be non-zero.
On the other hand, if $\mathrm{dim}(X) \ge 8$,
then the above anomaly may not vanish,
in which case one has to turn on background $G$-flux.
This typically happens in $M$-theory on 8-manifolds of
$Spin(7)$ holonomy and leads to interesting physics \cite{GS}.

Another, closely related condition that in general requires the $G$-flux
to be non-zero in compactification on an 8-manifold $X$
is the global tadpole anomaly \cite{SVW}:
\be
{\chi (X) \over 24} = N_{M2}
+ {1\over 2} \int_X {G \wedge G\over (2\pi)^2}
\label{sethrel}
\ee
Here, $\chi(X)$ is the Euler number of $X$ and $N_{M2}$
is the number of space-filling membranes. Clearly, this anomaly
is trivialised if the dimension of $X$ is less than eight.
We conclude that, in general, background $G$-flux
is required in compactifications of $M$-theory
on manifolds $X$ of dimension 8 (or greater). For instance, if we
consider vacua without membranes and 8-manifolds with non-zero
Euler number then non-zero $G$-flux is required.
Therefore, it is important to study which such compactifications
can be supersymmetric and, if so, what the corresponding
supersymmetry conditions are.

In this note we consider manifolds $X$ with metric
$g_X$ whose
holonomy group is $Spin(7)$ (or a subgroup theorof).
We obtain an $\mathcal{N}=1$
supersymmetric theory when $Hol({g_X}) = Spin(7)$.
We find that
the background flux  generates an effective superpotential
of the following simple form, originally proposed in \cite{sergei2}:
\be
W = \int_X G \wedge {\Omega}
\label{generalw}
\ee

Here $\Omega$ is the self-dual closed $Spin(7)$-invariant 4-form which
exists on any manifold of $Spin(7)$-holonomy.

Assuming that the typical size of $X$ is much larger than the Planck length $l_{Pl}$,
in the rest of this letter we show a complete agreement between
supersymmetry conditions in the eleven-dimensional supergravity
and in the effective three-dimensional theory with superpotential $W$.
We find that only particular choices of $G$-flux - characterised by
a particular representation of $Spin(7)$ are allowed - if we require
that the resulting
theory does not break supersymmetry spontaneously to leading order.
The conditions for unbroken supersymmetry in supergravity have previously
been studied in \cite{hawking,becker7}.

In addition to the potential for scalar
fields, we show that the abelian gauge fields in three dimensions
which originate from the $C$-field
in eleven dimensions can gain a mass due to  $G$-flux induced Chern-Simons couplings.
We also briefly comment on corrections to the leading potential including
those due to membrane and fivebrane instantons.

To conclude the introduction we will describe some elementary aspects of
the cohomology of $Spin(7)$ manifolds which we will require in our analysis of
supersymmetric vacua.
For more details
on the geometry of special holonomy manifolds we recommend
 \cite{joyce}.

\subsection{Cohomology of $Spin(7)$ Manifolds}

On a Riemannian manifold $X$, whose metric $g$ has holonomy
$H$, all fields ({\it i.e.} vectors, $p$-forms, spinors)
on $X$ form representations of $H$. With particular regard to $p$-forms on
$X$ this decomposition of forms commutes with the Laplacian and hence
the cohomology groups of $X$ are arranged into representations
of $H$. For example, the Hodge-Dolbeaut cohomology groups
$H^{p,q}(X, {\mathbb R})$
of a Kahler manifold $X$ consist of harmonic forms on
$X$ in a particular representation of $H$ $=$ $U({d \over 2})$.

For $X$ a manifold of $Spin(7)$-holonomy we obtain the following decompositions of
$H^k (X, \mathbb{R})$ which are induced from the decomposition of ${\Lambda}^k (\mathbb{R}^8)$
into irreducible representations of $Spin(7)$:
\begin{eqnarray}
H^0 (X, \mathbb{R}) & = & \mathbb{R} \nonumber \\
H^1 (X, \mathbb{R}) & = & H^1_{{\bf 8}} (X, \mathbb{R}) \nonumber \\
H^2 (X, \mathbb{R}) & = & H^2_{{\bf 7}} (X, \mathbb{R})
\oplus H^2_{{\bf 21}} (X, \mathbb{R}) \nonumber \\
H^3 (X, \mathbb{R}) & = & H^3_{{\bf 8}} (X, \mathbb{R})
\oplus H^3_{{\bf 48}} (X, \mathbb{R}) \nonumber \\
H^4 (X, \mathbb{R}) & = & H^4_{{\bf 1}^+} (X, \mathbb{R}) \oplus
H^4_{{\bf 7}^+} (X, \mathbb{R}) \oplus
H^4_{{\bf 27}^+} (X, \mathbb{R}) \oplus
H^4_{{\bf 35}^-} (X, \mathbb{R}) \label{hunderspin} \\
H^5 (X, \mathbb{R}) & = & H^5_{{\bf 8}} (X, \mathbb{R})
\oplus H^5_{{\bf 48}} (X, \mathbb{R}) \nonumber \\
H^6 (X, \mathbb{R}) & = & H^6_{{\bf 7}} (X, \mathbb{R})
\oplus H^6_{{\bf 21}} (X, \mathbb{R}) \nonumber \\
H^7 (X, \mathbb{R}) & = & H^7_{{\bf 8}} (X, \mathbb{R}) \nonumber \\
H^8 (X, \mathbb{R}) & = & \mathbb{R} \nonumber
\end{eqnarray}
The additional label ``$\pm$'' 
denotes self-dual/anti-self-dual four-forms, respectively.
The cohomology class of the 4-form
$\Omega$ generates $H^4_{{\bf 1}^+} (X, \mathbb{R})$.
We will denote the dimension
of $H^k_{\bf r} (X, \mathbb{R})$ as $b^k_{\bf r}$.

Thus far, we have only used the $Spin(7)$-structure locally. The fact that the metric
on $X$ has $Spin(7)$-holonomy implies global constraints on $X$ and this forces
some of the above groups to vanish when $X$ is compact. It will prove crucial to determine
which ones.

The reason we are interested
in $Spin(7)$-manifolds
in $M$-theory is that they
admit one covariantly constant (or parallel) spinor. This is the condition
for minimal supersymmetry in three dimensions in the absence of $G$-flux.
Since the metric on $X$ has $Spin(7)$ holonomy, it is Ricci flat and so
\be
D^2 = {\nabla}^2
\ee
where $D$ is the Dirac operator and $\nabla$
the covariant derivative.
Therefore, a zero mode of the Dirac operator is necessarily
a constant spinor and vice-versa.
Thus, we learn that the kernel of the Dirac operator on a manifold
of $Spin(7)$ holonomy is one dimensional
\footnote{Of course, on manifolds such as
Calabi-Yau fourfolds which are
$Spin(7)$ manifolds whose holonomy is a proper
subgroup of $Spin(7)$ there are more zero modes.}.
In fact, the index of the Dirac operator on a manifold with exactly
$Spin(7)$ holonomy
is precisely one \cite{joyce}.
This follows from the equation
above and the fact that manifolds of $Spin(7)$
holonomy have a constant spinor of only one chirality.
Therefore, on a manifold of $Spin(7)$ holonomy the cokernel of the
Dirac operator is empty.
Now, we will use the fact that spinors of any chirality on
a manifold of $Spin(7)$-holonomy can
actually be identified with certain combinations of
$p$-forms -- a fact which follows
essentially from ${\bf 8_s} \rightarrow {\bf 1} + {\bf 7}$
and ${\bf 8_c} \rightarrow {\bf 8}$ when $SO(8) \rightarrow Spin(7)$.

Namely, if $S = S_+ \oplus S_-$ is a spin bundle on $X$,
we have a natural isomorphism \cite{joyce}:
\be
S_+ \cong \Lambda^0_{{\bf 1}} \oplus \Lambda^2_{{\bf 7}},
\quad S_- \cong \Lambda^1_{{\bf 8}}
\ee

Furthermore, one can identify the Dirac operator
$D \colon C^{\infty} (S_+) \to C^{\infty} (S_-)$
with the following operator acting on differential forms:
\be
\pi_8 \circ d \colon
C^{\infty} (\Lambda^0_{{\bf 1}} \oplus \Lambda^2_{{\bf 7}})
\to C^{\infty} (\Lambda^1_{{\bf 8}})
\ee
{\it i.e.}
we take the exterior derivative and project the result onto the
eight-dimensional representation of $Spin(7)$.

Therefore, the Dirac index (also called the $A$-roof genus)
on the compact $Spin(7)$
manifold $X$ can be written:
\be
\hat A (X) = b^0_{\bf 1} + b^2_{{\bf 7}} - b^1_{{\bf 8}} =
1 + b^2_{{\bf 7}} - b^1_{{\bf 8}}
\ee

In particular, if $Hol(g_X)=Spin(7)$,
$X$ is simply-connected and as we saw above has $\hat A(X) = 1$.
Therefore, we have $b^1_{{\bf 8}}=0$ and $b^2_{{\bf 7}}=0$.
Using the canonical isomorphisms
(which are easily obtained by wedging and
contracting with $\Omega$) \cite{joyce}:
\be
\Lambda^1_{{\bf 8}} \cong
\Lambda^3_{{\bf 8}} \cong
\Lambda^5_{{\bf 8}} \cong
\Lambda^7_{{\bf 8}}, \quad
\Lambda^2_{{\bf 7}} \cong
\Lambda^4_{{\bf 7}} \cong
\Lambda^6_{{\bf 7}}
\ee
we obtain further constraints
$b^1_{{\bf 8}} = b^3_{{\bf 8}} = b^5_{{\bf 8}} = b^7_{{\bf 8}}=0$
and $b^2_{{\bf 7}} = b^4_{{\bf 7}} = b^6_{{\bf 7}}=0$.

To summarize, if $X$ is a compact 8-manifold, such that
$\mathrm{Hol} (g_X) = Spin(7)$, then the cohomology of $X$ can be
decomposed into the following representations of $Spin(7)$:
\begin{eqnarray}
H^0 (X, \mathbb{R}) & = & \mathbb{R} \nonumber \\
H^1 (X, \mathbb{R}) & = & 0 \nonumber \\
H^2 (X, \mathbb{R}) & = & H^2_{{\bf 21}} (X, \mathbb{R}) \nonumber \\
H^3 (X, \mathbb{R}) & = & H^3_{{\bf 48}} (X, \mathbb{R}) \nonumber \\
H^4 (X, \mathbb{R}) & = & H^4_{{\bf 1}^+} (X, \mathbb{R}) \oplus
H^4_{{\bf 27}^+} (X, \mathbb{R}) \oplus
H^4_{{\bf 35}^-} (X, \mathbb{R}) \label{hunderspin7} \\
H^5 (X, \mathbb{R}) & = & H^5_{{\bf 48}} (X, \mathbb{R}) \nonumber \\
H^6 (X, \mathbb{R}) & = & H^6_{{\bf 21}} (X, \mathbb{R}) \nonumber \\
H^7 (X, \mathbb{R}) & = & 0 \nonumber \\
H^8 (X, \mathbb{R}) & = & \mathbb{R} \nonumber
\end{eqnarray}

In this list, the largest representation structure
appears in degree 4.
Since we are going to consider $M$-theory backgrounds
with non-trivial 4-form flux $G$, this cohomology
group also plays an important role in our discussion.
In particular, it will be crucial that on a compact
manifold $X$ of exactly $Spin(7)$ holonomy we have
$H^4_{{\bf 7}^+} (X, \mathbb{R}) =0$.

\sect{Supersymmetry Conditions in $D=11$ Supergravity}

Now we consider the conditions for unbroken supersymmetry in
eleven-dimensional supergravity on a $Spin(7)$ manifold $X$.
The supergravity approximation to $M$-theory is valid as long as
the size of $X$ is large, compared to the Planck scale.
Supersymmetry conditions in (warped) compactifications of $M$-theory
to three dimensional Minkowski space-time have already been discussed
in the literature \cite{becker2} and in fact the conditions for unbroken
supersymmetry upon compactification on a $Spin(7)$ manifold have
also been obtained \cite{hawking,becker7,gary7}.
Also, the solutions to the equations of
motion of $M$-theory on
Kahler 8-manifolds have been discussed in \cite{becker8}.
In this section, we slightly extend the analysis of supersymmetric vacua,
allowing for the possibility that the three-dimensional cosmological
constant is non-zero. In other words, we assume
the eleven-dimensional space-time to be of the form:
$$
M^3 \times X
$$
where $M^3$ is a maximally symmetric three-dimensional space.
More precisely, we consider a warped product of $M^3$ and $X$,
rather than a direct product. If we denote the scalar warp
factor $\Delta (y^m)$, then the corresponding metric reads:
\be
ds^2 = e^{2 \Delta /3} \eta_{\mu \nu} (M^3)~ dx^{\mu} dx^{\nu}
+ e^{-\Delta /3} g_{mn}(X)~ dy^m dy^n 
\label{metric}
\ee
The external components of the 3-form field $C$ have the form:
\be
C_{012} = - e^{\Delta}
\ee
Finally, we put no restrictions on the internal components
of the $G$-flux.

Since all fermionic fields vanish in the background, we can
focus only on the supersymmetry variation of the gravitino field:
\be
\delta \psi_M = \nabla_M \eta
- {1 \over 288} G_{PQRS} (\Gamma_M^{PQRS} - 8 \delta_M^P \Gamma^{QRS}) \eta
\label{delpsi}
\ee
where $\eta$ is a supersymmetry variation parameter.

Now we require $\delta \psi_M = 0$ and consider different
components of this equation. Since the calculation is
pretty standard (see {\it e.g.} \cite{becker2}),
here we only outline the main steps. First, one makes
the $3+8$ split, compatible with the metric (\ref{metric}):
\be
\Gamma_{\mu} = e^{\Delta /3} (\gamma_{\mu} \otimes \gamma_9),
\quad
\Gamma_{m} = e^{-\Delta /6} ({\bf 1} \otimes \gamma_m)
\ee
Similarly, one can decompose the supersymmetry
parameter $\eta$ into an eight-dimensional spinor $\xi$
on $X$ (such that $\xi^T \xi =1$ and $\gamma_9 \xi = + \xi$)
and into a three-dimensional spinor $\epsilon$ on $M^3$,
which obeys\footnote{
In the three-dimensional effective supergravity theory,
$m_{\psi}$ has interpretation as a gravitino mass parameter,
which is related to the cosmological constant in the usual way.}
$\nabla_{\mu} \epsilon = m_{\psi} {\gamma}_{\mu} \epsilon$.
Specifically, we have:
\be
\eta = e^{\Delta /6} (\epsilon \otimes \xi)
\ee
After rescaling transformations that eliminate
the dependence on the warp factor $\Delta$, from
the internal components of the supersymmetry
variation (\ref{delpsi}) we obtain the following
supersymmetry condition:
\be
m_{\psi} \gamma_m \xi - {1 \over 12} G_{mpqr} \gamma^{pqr} \xi = 0
\ee
However, it turns out to be compatible with
the external components
of (\ref{delpsi}) if and only if $m_{\psi}=0$,
{\it i.e.} when three-dimensional space-time is flat.
Hence, the supersymmetry conditions take
the form obtained earlier in \cite{becker2}:
$$
G_{mpqr} \gamma^{pqr} \xi = 0
$$

If we multiply this relation by $\gamma^n$ and
by $\xi^T$ from the left and use the identity:
\be
\Omega_{mnpq} = \xi^T \gamma_{mnpq} \xi
\ee
we can express this supersymmetry condition
in terms of the Cayley 4-form $\Omega$:
\be
G_{mpqr} \Omega_n^{\;\;pqr} =0
\label{susycond}
\ee
It is convenient to denote the left-hand side of this
equation as $T_{mn} = G_{mpqr} \Omega_n^{\;\;pqr}$.
Then, the above supersymmetry condition reads:
\be
T_{mn}=0
\ee

Let us analyze different components of these equations.
$T_{mn}$ is a 2-index tensor field on $X$. Since $g_X$ has $Spin(7)$
holonomy, we can consider decomposing $T$ into irreducible $Spin(7)$
representations. Which representations appear? If $g_X$ had generic,
{\it i.e.} $SO(8)$ holonomy, then $T$ decomposes into traceless symmetric,
antisymmetric, and trace components. As $SO(8)$ representations these
have dimensions ${\bf 35, 28}$ and ${\bf 1}$ respectively. But as $Spin(7)$
representations, the ${\bf 35}$ and ${\bf 1}$ remain irreducible, whilst
${\bf 28}$ becomes ${\bf 7} + {\bf 21}$.

Now, $T$ is not an arbitrary 2-tensor, but
a tensor constructed from two 4-forms $G$ and $\Omega$.
The fact that $\Omega$
is in the trivial representation of $Spin(7)$
implies that the representations in which
$T$ resides can at most be those of $G$.
Then, the fact that 4-forms on a $Spin(7)$
manifold can only be in the representations ${\bf 1}$,
${\bf 7}$, ${\bf 27}$ or ${\bf 35}$
implies that the antisymmetric
part of $T$ cannot contain any component in
the representation ${\bf 21}$. Therefore, we learn that the
condition $\mathrm{tr} T = 0$ means that $G$
is not a $Spin(7)$ singlet. In other words $G_{{\bf 1}^+}=0$.
The condition that the symmetric part of $T_{mn}$
vanishes implies that $G$ is self-dual, {\it i.e.}
the ${{\bf 35}}$ piece of $G$ must vanish.
Finally, the condition that the antisymmetric part of
$T_{mn}$ vanishes says that the ${{\bf 7}}$ piece of $G$ vanishes.
Therefore, according to (\ref{hunderspin7}),
the 4-form field $G$ compatible with $\mathcal{N}=1$
supersymmetry can have non-vanishing components
only in the ${{\bf 27}}$ representation of $Spin(7)$:
\be
G \in H^4_{{\bf 27}^+} (X, \mathbb{R})
\label{susycohom}
\ee

Before we proceed to the interpretation of this supersymmetry
condition in the effective $\mathcal{N}=1$ three-dimensional
theory, let us remark that since it is derived as a
local condition on the $G$-field
the result is valid even when $X$ is a
non-compact $Spin(7)$ manifold. Such manifolds usually
appear as local models in the study of $Spin(7)$
singularities \cite{GS} and play an important role in
the geometric engineering of $\mathcal{N}=1$ three-dimensional
gauge theories decoupled from gravity.


\sect{Interpretation in the Effective Three-Dimensional Theory}

In this section we will interpret the above results
in terms of the effective $\mathcal{N}=1$ three-dimensional theory.
What is the effective three dimensional theory? When $X$ is large,
and $G$ is zero, standard Kaluza-Klein analysis applies and it is
straightforward to see \cite{paptown} that the three dimensional
low energy theory
is ${\cal N}=1$ supergravity with $b^2_{\bf 21}$ vector multiplets $A_i$
(whose
vectors arise from the 3-form potential $C$). There are also $b^3_{\bf 48}$
scalar multiplets $\rho_j$ from the 3-form and $b^4_{\bf 35} + 1$ scalar
multiplets $\phi_k$ from the metric tensor. The latter fields parametrise
locally the space of $Spin(7)$ holonomy metrics on $X$ which are near $g_X$.
The gauge group is locally $U(1)^{b^2_{\bf 21}}$ but globally
$H^{2}(X, U(1))$. As we discussed in the introdution, however, the theory
without $G$-flux may not be a consistent $M$-theory vacuum. We can regard
the theory with $G$-flux as adding extra couplings to the above theory
in which the $A,\rho$ and $\phi$ fields are massless and non-interacting to
leading order in $l_{pl}$.

The small amount of supersymmetry allows for a rich dynamical
structure in these theories, and a variety of interaction
terms in the effective Lagrangian. For this reason,
it is convenient to write the effective Lagrangian in superspace,
which makes $\mathcal{N}=1$ supersymmetry manifest.
Minimal three-dimensional superspace can be obtained\footnote{\cite{susybook} is a
useful reference.}
by combining three-dimensional coordinates $x^{\mu}$
with real Grassmann variables $\theta_{\alpha}$,
and by introducing the corresponding covariant derivatives $D_{\alpha}$.
Then, the effective three-dimensional Lagrangian can be
schematically written as a full superspace intagral:
\be
L_{3D} = \int d^3 x d^2 \theta E^{-1} K
+ \int d^3 x d^2 \theta E^{-1} W (\rho_j, \phi_k)
\label{leffective}
\ee
where the first term represents the kinetic action,
while $W$ (unlike $K$) depends only on the scalar fields
but not their derivatives.
After we perform $d^2 \theta$ integral in (\ref{leffective}),
the last term leads to the scalar potential in the effective
theory \cite{susybook}.

In a supersymmetric vacuum with zero cosmological
constant, the following conditions must be satisfied:
\be
W=0, \quad \quad {\partial W \over \partial \rho_j}={\partial W \over
\partial \phi_k} = 0
\label{susydw}
\ee
These are the supersymmetry conditions that we want to compare
to the ones in eleven-dimensional supergravity.

Following \cite{sergei2,AS}, we interpret
the supersymmetry condition (\ref{susycohom}) in terms of
the effective superpotential $W$ induced by the $G$-flux:
\be
W = {1 \over 2 \pi} \int_X G \wedge \Phi
\label{wforg}
\ee
In general, the expression (\ref{generalw}) for the effective
superpotential was conjectured from the identification of
BPS domain walls with branes wrapped on supersymmetric
submanifold $S \subset X$. In our case, these are M5-branes
wrapped on Cayley 4-cycles, with tension:
\be
T \ge \int_S \Phi = \vert \Delta W \vert
\label{bpsbound}
\ee
Here, we will justify the formula (\ref{wforg}) for the effective
superpotential by showing that the eleven-dimensional
supersymmetry conditions (\ref{susycohom}) and the corresponding
conditions (\ref{susydw}) in the effective three-dimensional theory are the same.

The first equation in (\ref{susydw}), namely $W=0$, implies
that three-dimensional cosmological constant is zero, and from
(\ref{wforg}) (which is proportional to $trT_{mn}$ from the
previous section)
we find that it requires the singlet piece of $G$ to vanish.

On the other hand, $\partial W / \partial \phi_i$
is the variation of $W$ with respect to the scalar
fields which come from the metric deformations of the
compact $Spin(7)$ holonomy manifold $X$. The superpotential
$W$ can only depend on these scalars since it only depends on $\Omega$.
According to \cite{joyce},
the latter generate $H^4_{{\bf 35}^-} (X, \mathbb{R})$\footnote{In other words, if we
add any small harmonic anti-self-dual 4-form to $\Omega$ we get a new
$Spin(7)$ structure and a correspondingly new $Spin(7)$ holonomy metric.},
so that the second equation in (\ref{susydw}) implies
$G_{{\bf 35}^-}=0$.
Since for a compact $Spin(7)$ manifold
$H^4_{{\bf 7}^+} (X, \mathbb{R})=0$,
we conclude that $G$-flux has to be an element of
$H^4_{{\bf 27}^+} (X, \mathbb{R})$, in complete agreement
with the supergravity result (\ref{susycohom}):
\be
G \in H^4_{{\bf 27}^+} (X, \mathbb{R})
\ee



\subsection{Quantum Corrections to the Potential}

The expression (\ref{generalw})
represents the leading contribution to the potential induced
by the non-trivial $G$-flux. In this section we will briefly
discuss the ``perturbative'' and non-perturbative contributions
to the potential.

The total superpotential schematically can be written as:
\be
W_{\mathrm{tot}} = W + W_{\mathrm{pert}} + W_{\mathrm{non-pert}}
\ee
where $W$ is the classical term (\ref{wforg}).
We will first discuss the perturbative
contributions and then the non-perturbative ones.
The one-loop contribution to the perturbative superpotential
$W_{\mathrm{pert}}$ is expected to be in the following simple form:
\be
W_{\mathrm{pert}} = {1 \over 4 \pi} G^{ab}
{\partial^2 W (\phi) \over \partial \phi^a \partial \phi^b}
= {1 \over 8 \pi^2} \int_X G \wedge \delta^2 \Phi + \ldots
\label{ftwpert}
\ee
where $G^{ab}$ is a scalar field metric. This follows essentially from
the supersymmetry of the theory.
The best way to demonstrate this is to compactify
the three-dimensional theory further on a circle.
This leads to a supersymmetric field theory in two dimensions,
which can be also thought of as a result of Type IIA compactification on $X$.
The `classical' superpotential in this theory also has the form
(\ref{wforg}), whereas (\ref{ftwpert}) is a one-loop anomaly
\cite{Shifman}. In fact, \cite{Shifman} argue that there are no additional
contributions at higher loop order.

The non-perturbative part of the superpotential is generated
by $M2$-brane and $M5$-brane instantons wrapped on
three and six-cycles, $V^3$ and $V^6$, respectively. Since such cycles
are not supersymmetric, such instantons break both supersymmetries.
Consequently, the two corresponding Goldstone fermions imply that these
wrapped branes will contribute to $W$. The form of these contributions is
of the form
\be
W_{\mathrm{non-pert}} \sim
\sum_{V^3} e^{- \mathrm{Vol} (V^3) - \int_{V^3}C} +
\sum_{V^6} e^{- \mathrm{Vol} (V^6) - \int_{V^6}\hat{C}}
\ee

Here $C$ is the three-form field and $\hat{C}$ is its dual. Note that
the period of $C$ through $V^3$ is a function of the scalars $\rho_j$
and that the period of $\hat{C}$ is formally a function of the scalars
which are dual to the photon fields.
It is conceivable that for generic $G$-flux
the total superpotential $W_{tot}$
in the $\mathcal{N}=1$ effective three-dimensional theory
has only isolated fixed points.

Finally, we remark that the $U(1)^{b^2}$ gauge fields are
also typically massive in the presence of $G$-flux due to
Chern-Simons couplings \cite{GS}.
Inserting the Kaluza-Klein ansatz for the $C$-field,
\be
C = \Sigma_I \alpha_I \wedge A^I (x) + ..
\ee
into the interaction,
\be
\int C \wedge G \wedge G
\ee
gives rise to the following three-dimensional Chern-Simons action for the
three-dimensional gauge fields $A^I$
\be
C_{IJ}\int A^I \wedge dA^J
\ee
where
\be
C_{IJ} = \int_X \alpha_I \wedge \alpha_J \wedge G
\ee
where $G$ is the background $G$-flux. For generic enough $G$ the couplings
$C_{IJ}$ will be non-zero and therefore all the gauge fields gain a mass.
For instance, if we take $G$ $=$ $\Omega$ then all the diagonal couplings
$C_{II}$ are non-zero and are in fact given by
\be
C_{II} = -2\int_X \alpha_I \wedge * \alpha_I
\ee
This follows from the fact that the 2-forms $\alpha_I$ are all in the
${\bf 21}$-dimensional representation of $Spin(7)$ and as such they satisfy
\be
- \alpha_I \wedge \Omega = 2 *\alpha_I
\ee

\medskip

\centerline{\bf Acknowledgments}
\noindent

We would like to thank P. Candelas, D. Joyce, J.~Sparks, and E.~Witten
for useful discussions.
This research was partially conducted during the period S.G.
served as a Clay Mathematics Institute Long-Term Prize Fellow.
The work of S.G. is also supported in part by grant RFBR No. 01-02-17488,
and the Russian President's grant No. 00-15-99296.


\end{document}